\documentclass[aps,prx,twocolumn,superscriptaddress,showpacs,longbibliography]{revtex4-2}
\usepackage{amsmath,amssymb,amsfonts,amsthm}
\usepackage{graphicx}
\usepackage{bm}
\usepackage{bbm}
\usepackage{bbold}
\usepackage{mathrsfs}
\usepackage{color}
\usepackage{dcolumn}   
\usepackage{epstopdf}
\usepackage[caption=false]{subfig}
\usepackage{xr}
\usepackage[colorlinks=true,linkcolor=blue,urlcolor=blue,citecolor=blue]{hyperref}

\begin{document}

 \newcommand{\breite}{1.0} 

\newtheorem{prop}{Proposition}
\newtheorem{cor}{Corollary} 

\newcommand{\be}{\begin{equation}}
\newcommand{\ee}{\end{equation}}

\newcommand{\bea}{\begin{eqnarray}}
\newcommand{\eea}{\end{eqnarray}}
\newcommand{\lt}{<}
\newcommand{\gt}{>} 

\newcommand{\Reals}{\mathbb{R}}     
\newcommand{\Com}{\mathbb{C}}       
\newcommand{\Nat}{\mathbb{N}}       

\newcommand{\id}{\mathbboldsymbol{1}}    

\newcommand{\Real}{\mathop{\mathrm{Re}}}
\newcommand{\Imag}{\mathop{\mathrm{Im}}}

\def\O{\mbox{$\mathcal{O}$}}   
\def\F{\mathcal{F}}			
\def\sgn{\text{sgn}}

\newcommand{\deo}{\ensuremath{\Delta_0}}
\newcommand{\dea}{\ensuremath{\Delta}}
\newcommand{\ak}{\ensuremath{a_k}}
\newcommand{\ad}{\ensuremath{a^{\dagger}_{-k}}}
\newcommand{\sx}{\ensuremath{\sigma_x}}
\newcommand{\sz}{\ensuremath{\sigma_z}}
\newcommand{\spl}{\ensuremath{\sigma_{+}}}
\newcommand{\smi}{\ensuremath{\sigma_{-}}}
\newcommand{\alk}{\ensuremath{\alpha_{k}}}
\newcommand{\bk}{\ensuremath{\beta_{k}}}
\newcommand{\ok}{\ensuremath{\omega_{k}}}
\newcommand{\vd}{\ensuremath{V^{\dagger}_1}}
\newcommand{\vi}{\ensuremath{V_1}}
\newcommand{\vo}{\ensuremath{V_o}}
\newcommand{\zc}{\ensuremath{\frac{E_z}{E}}}
\newcommand{\xc}{\ensuremath{\frac{\Delta}{E}}}
\newcommand{\xd}{\ensuremath{X^{\dagger}}}
\newcommand{\aok}{\ensuremath{\frac{\alk}{\ok}}}
\newcommand{\tpw}{\ensuremath{e^{i \ok s }}}
\newcommand{\tpe}{\ensuremath{e^{2iE s }}}
\newcommand{\tmw}{\ensuremath{e^{-i \ok s }}}
\newcommand{\tme}{\ensuremath{e^{-2iE s }}}
\newcommand{\epls}{\ensuremath{e^{F(s)}}}
\newcommand{\emis}{\ensuremath{e^{-F(s)}}}
\newcommand{\epl}{\ensuremath{e^{F(0)}}}
\newcommand{\emi}{\ensuremath{e^{F(0)}}}

\newcommand{\lr}[1]{\left( #1 \right)}
\newcommand{\lrs}[1]{\left( #1 \right)^2}
\newcommand{\lrb}[1]{\left< #1\right>}
\newcommand{\nbt}{\ensuremath{\lr{ \lr{n_k + 1} \tmw + n_k \tpw  }}}

\newcommand{\om}{\ensuremath{\omega}}
\newcommand{\dw}{\ensuremath{\Delta_0}}
\newcommand{\wbp}{\ensuremath{\omega_0}}
\newcommand{\dv}{\ensuremath{\Delta_0}}
\newcommand{\vbp}{\ensuremath{\nu_0}}
\newcommand{\vplus}{\ensuremath{\nu_{+}}}
\newcommand{\vminus}{\ensuremath{\nu_{-}}}
\newcommand{\wplus}{\ensuremath{\omega_{+}}}
\newcommand{\wminus}{\ensuremath{\omega_{-}}}
\newcommand{\uv}[1]{\ensuremath{\mathbf{\hat{#1}}}} 
\newcommand{\abs}[1]{\left| #1 \right|} 
\newcommand{\norm}[1]{\left \lVert #1 \right \rVert} 
\newcommand{\avg}[1]{\left< #1 \right>} 
\let\underdot=\d 
\renewcommand{\d}[2]{\frac{d #1}{d #2}} 
\newcommand{\dd}[2]{\frac{d^2 #1}{d #2^2}} 
\newcommand{\pd}[2]{\frac{\partial #1}{\partial #2}} 
\newcommand{\pdd}[2]{\frac{\partial^2 #1}{\partial #2^2}} 
\newcommand{\pdc}[3]{\left( \frac{\partial #1}{\partial #2}
 \right)_{#3}} 
\newcommand{\ket}[1]{\left| #1 \right>} 
\newcommand{\bra}[1]{\left< #1 \right|} 
\newcommand{\braket}[2]{\left< #1 \vphantom{#2} \right|
 \left. #2 \vphantom{#1} \right>} 
\newcommand{\matrixel}[3]{\left< #1 \vphantom{#2#3} \right|
 #2 \left| #3 \vphantom{#1#2} \right>} 
\newcommand{\grad}[1]{{\nabla} {#1}} 
\let\divsymb=\div 
\renewcommand{\div}[1]{{\nabla} \cdot \boldsymbol{#1}} 
\newcommand{\curl}[1]{{\nabla} \times \boldsymbol{#1}} 
\newcommand{\laplace}[1]{\nabla^2 \boldsymbol{#1}}
\newcommand{\vs}[1]{\boldsymbol{#1}}
\let\baraccent=\= 
\newcommand{\ka}[1]{{\color{magenta} #1}}


\title{Nanoscale defects as probes of time reversal symmetry breaking}

\author{Suman Jyoti De}
\affiliation{Physics Department, McGill University, Montr\'eal, QC, Canada H3A 2T8}
\author{T. Pereg-Barnea}
\affiliation{Physics Department, McGill University, Montr\'eal, QC, Canada H3A 2T8}
\affiliation{ICFO–Institut de Ci\'encies Fot\'oniques, The Barcelona Institute of Science and Technology, Castelldefels, Barcelona, Spain 08860}
\author{Kartiek Agarwal}
\affiliation{Material Science Division, Argonne National Laboratory, Lemont, IL, USA 60548}
\affiliation{Physics Department, McGill University, Montr\'eal, QC, Canada H3A 2T8}

\date{\today}
\begin{abstract}
Nanoscale defects such as Nitrogen Vacancy (NV) centers can serve as sensitive and non-invasive probes of electromagnetic fields and fluctuations from materials, which in turn can be used to characterize these systems. Here we specifically discuss how NV centers can probe time-reversal symmetry breaking (TRSB) phenomena in low-dimensional conductors. We argue that the difference in relaxation rates $\Gamma_{\pm \hat{z}}$ of NV centers starting from $m = \pm 1$ spin states to the ground state with $m = 0$ directly probes TRSB. The effect arises from the difference in the fluctuation spectrum of left and right-polarized electromagnetic fields emanating from such materials. In the quantum Hall setting, the NV center experiences (nearly zero) large additional contribution to its relaxation due to the presence of the material when its magnetic dipole (anti-) aligns with the external field. More generally, the difference in the relaxation rates is sensitive to the imaginary part of the wave-vector dependent Hall conductivity. We argue that this can be used to determine the Hall viscosity, which can potentially distinguish candidate fractional quantum Hall states and be used to infer the pairing angular momentum in TRSB superconductors. For the latter, we consider specifically the case of TRSB in stacked twisted Bismuth strontium calcium copper oxide (BSCCO) flakes, which have recently been investigated experimentally and are suggested to exhibit TRSB. We show that the average relaxation rate $\left[\Gamma_{+\hat{z}} + \Gamma_{-\hat{z}}\right]$ near such a system exhibits a Hebel-Slichter like enhancement below $T_c$. The difference $\Gamma_{+\hat{z}} - \Gamma_{-\hat{z}}$ also inherits this peak but is only non-zero for $T < T_c$ and only in a chiral d-wave superconductor. We provide concrete estimates for observing this effect. 
\end{abstract}
\maketitle

\section{Introduction}

Nanoscale defects such as Nitrogen Vacancy (NV) centers have emerged as a novel probe of magnetic and electronic correlations in materials~\cite{maze2008nanoscale,rovny2024new,marchiori2022nanoscale,abobeih2019atomic}. NV centers in particular behave as qubits with a magnetic dipole moment and intrinsic level splitting~\cite{PhysRevB.74.161203}. When the NV center is placed near a material, it experiences a different electromagnetic environment to what it experiences in the absence of the sample. Static fields can be mapped out by changes to the level splitting(s) of the NV center states~\cite{maze2008nanoscale} (similarly to spin resonance measurements), while magnetic fluctuations orthogonal to its intrinsic dipole axis result in relaxation of the qubit at a rate proportional to the strength of magnetic fluctuations at frequency equal to the intrinsic splitting~\cite{kolkowitz2015probing,agarwal2017magnetic,ariyaratne2018nanoscale}. These magnetic fields are ideally dominated by current and/or spin fluctuations in the probed material, and thus contain information about the conductivity or spin susceptibility of the material. 

Two aspects of such defects sets them apart from traditional probes. First, they are non-invasive since they do not require the application of external fields---one can view them analogously to lab microscopes that passively gather light from a specimen of interest. Second, being atomic scale point-defects, they are  sensitive to highly local material properties. For instance, as discussed in Ref.~\cite{agarwal2017magnetic}, these probes can be used to infer wave-vector dependent conductivity in materials, at wave-vector $q \sim 1/z_{\text{NV}}$, where $z_{\text{NV}}$ is the distance of the probe from the material. (In comparison, Nuclear Magnetic Resonance which is based on similar principles, naturally probes magnetic fluctuations at wave-vectors related to the crystal structure~\cite{slichter2013principles}.) This make these probes useful from the perspective of detecting novel correlations in materials where usual conductance or global spin-susceptibility measurements may not implicate new phenomena, or where multiple orders coexist and external fields can influence the state making measurements challenging to interpret. 

NV centers have been used to map out local spin patterns~\cite{appel2019nanomagnetism,finco2021imaging,haykal2020antiferromagnetic}, measure spin wave spectra in low dimensional magnets~\cite{lee2020nanoscale,simon2022filtering}, transport  characteristsics (ballistic/diffusive/hydrodynamic/superconducting) of electrons~\cite{kolkowitz2015probing,vool2021imaging,jenkins2022imaging,monge2023spin,thiel2016quantitative,schlussel2018wide,pelliccione2016scanned}, measure phononic Cherenkov effect in photoexcited graphene~\cite{andersen2019electron}, among others. Immense progress has been made in engineering their intrinsic relaxation properties~\cite{balasubramanian2009ultralong} to obtain $T_1$ times of the order of seconds at cryogenic temperatures~\cite{abobeih2018one}, attaching them on maneuverable tips~\cite{wan2018efficient,hedrich2020parabolic} and controlled implantation at various depths and in ensemble~\cite{aharonovich2012homoepitaxial,guo2021tunable,khanaliloo2015high,challier2018advanced}. Other nanoscale defects such as silicon vacancy centers~\cite{zuber2023shallow} among others~\cite{PhysRevB.105.224106} are also potential candidates for novel quantum sensing capabilities~\cite{wolfowicz2021quantum}.  Theoretically, proposals have been made to use NV centers to measure hydrodynamic transport and Kondo impurities in metals~\cite{agarwal2017magnetic}, crystallization at low densities in electron gases~\cite{dolgirev2023local}, spinon Fermi surfaces in spin liquid candidates~\cite{lee2023proposal}, and novel edge states in topological materials~\cite{rodriguez2018probing}, among others~\cite{PhysRevResearch.4.L012001}. 

In this work, we focus on how such probes can be used to ascertain time reversal symmetry breaking (TRSB) phenomena in low dimensional materials of various kinds, including quantum Hall systems, and TRSB superconductors. A key insight as to the utility of these probes comes from a simple observation---TRSB materials exhibit a circular dichroism in the electromagnetic fluctuation spectrum near them. Specifically, the spectrum of left and right circularly polarized magnetic fields is different near a TRSB material. This dichroism stems directly from the spectra of `circularly polarized' current and spin operators $j_\pm = j_x \pm i j_y, s_\pm = s_x \pm i s_y$ (defined analogously as in electromagnetism) that are inherently different in a TRSB material. The most stark example of this physics is the quantum Hall insulator---operators $j_+$ can be used to raise Landau level index from the ground state, while  $j_-$ reduces the Landau level index~\cite{tong2016lectures}. At low temperatures, the spectrum of these raising and lowering operators is thus significantly different from one another, and this translates to dichroism in magnetic fluctuations near the material. 

In Sec.~\ref{sec:dichorism}, we elaborate further on this dichroism, computing the $1/T^{\hat{z}}_1, 1/T^{-\hat{z}}_1$ relaxation rates of a magnetic dipole probe that is oriented either perpendicular, in the $\hat{z}$ direction, or anti-perpendicular, in the $-\hat{z}$ direction, to the surface of the material of interest. Note that for NV center based imagining, these are realized with a single NV center with its magnetic dipole oriented along the $\hat{z}$ axis, and measuring its relaxation to equilibrium starting from either the $\ket{m_s = 1}$ or $\ket{m_s = -1}$ states. In the first case ($m_s = 1$), the relaxation rate amounts to a sum of spectral weights $\abs{b_+(\omega_{\text{NV}})}^2 + \abs{b_- (-\omega_{\text{NV}})}^2$, while in the second case, one probes $\abs{b_+(-\omega_{\text{NV}})}^2 + \abs{b_- (\omega_{\text{NV}})}^2$, with $\omega_{\text{NV}}$ being the transition frequency of the NV center. Here $b_\pm = b_x + i b_y$ where $b_x, b_y$ are the $x,y$ components of the magnetic field on the defect site, respectively; see Eqs.~(\ref{eq:chiralrelaxation}) for the precise definition. As we show, the two components in these rates are related by detailed balance as one would expect of emission and absorption rates due to fluctuations in a thermally equilibrated system. However, these two rates consist of a cross correlation of $b_x, b_y$ fluctuations, specifically $\text{Im} \left[N_{xy}\right]$ [see Eqs.~(\ref{eq:T1})] that is odd under time-reversal. This piece is precisely isolated by considering the difference of the two relaxation rates, $1/T^{\hat{z}}_1 - 1/T^{-\hat{z}}_1$, and which directly implicates TRSB if it is non-zero. 
\begin{figure}
    \includegraphics[width=3in]{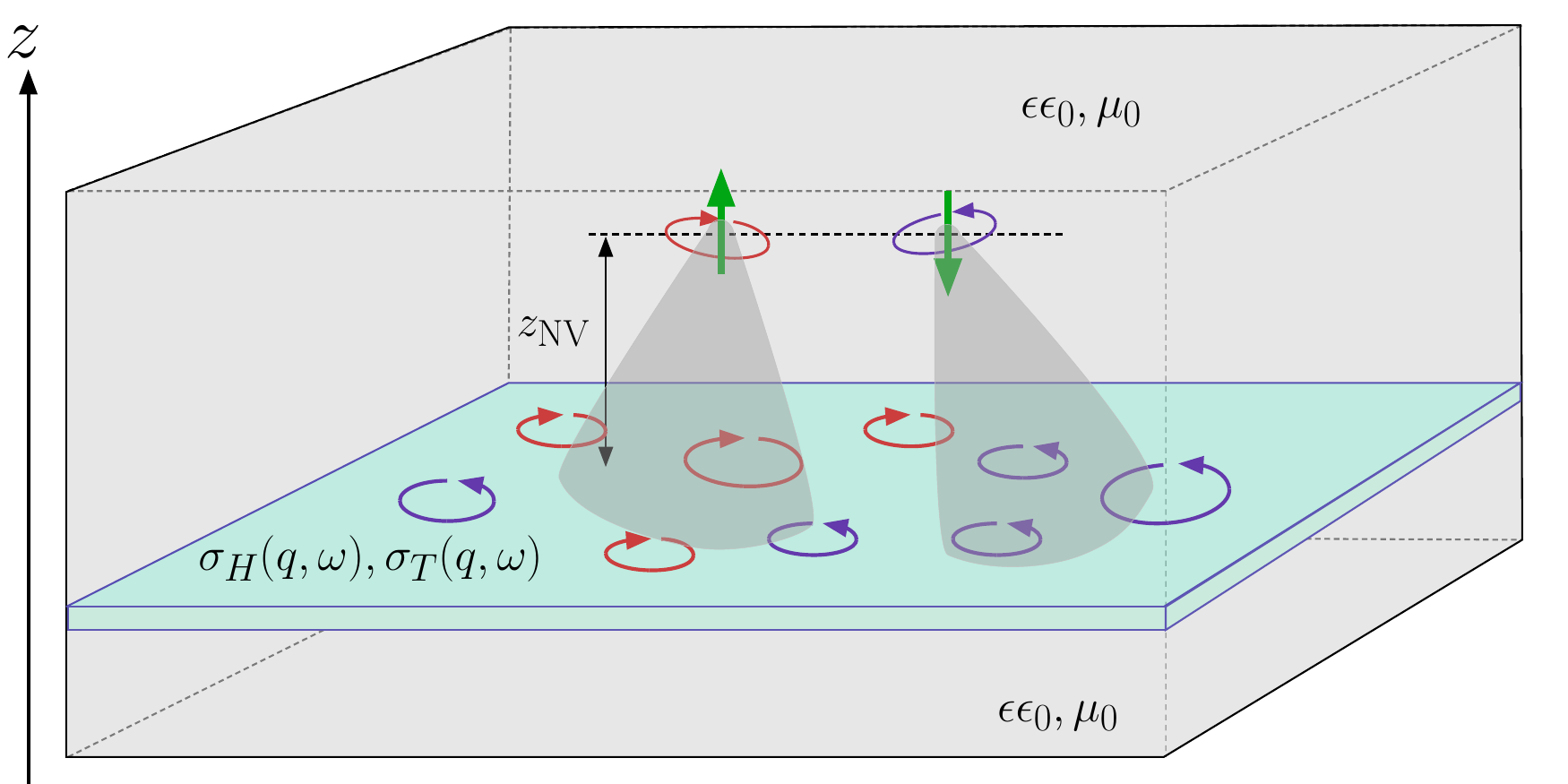}
    \caption{Magnetic noise from a time-reversal symmetry breaking sample (blue) exhibits a different spectrum for magnetic fluctuations with opposite chiralities (purple and red). The NV center is assumed to be embedded here in a dielectric of constant $\epsilon_0$. We assume for simplicity that the sample sits sandwiched between two dielectrics of similar dielectric constant. NV centers with magnetic dipoles (green arrows) oriented towards or against such a material exhibit different relaxation rates owing to this difference. (Note these are not implied to be two separate NV centers but the same color center initialized in the state $\ket{m_s = \pm 1}$. They are shown separately here for clarity.) In particular, $\Gamma^{\pm \hat{z}}_{\text{ems}} \propto \text{Re} \left[ \sigma_T \mp i \sigma_H \right] (q \sim 1/z_{\text{NV}}, \omega = \omega_{\text{NV}} $; see main text for details.}
    \label{fig:trsbnoise}
\end{figure}
In Sec.~\ref{sec:sigmaxy}, we make a connection between $\text{Im} \left[N_{xy}\right]$, and $\text{Im} \left[\sigma_{H} (q, \omega) \right]$, that is, the imaginary part of the Hall conductivity at finite wave-vectors in the material. In particular, we study the electromagnetic fluctuation spectrum near a two dimensional TRSB material sandwiched between dielectrics. Following the standard analysis of electromagnetic waves reflecting off of conducting surfaces~\cite{born2013principles}, we compute the magnetic response of the system in the presence of an added magnetic dipole, which generates both $s-$ and $p-$ polarized waves. We find that the susceptibility of interest, $\chi_{xy}$ depends only on reflections that convert $s-$ polarized light to $p-$ polarized light and vice versa, an effect fully absent in systems with time-reversal symmetry. We use the fluctuation dissipation theorem to arrive at our final result for $\text{Im} \left[N_{xy}\right]$. We find that much like the usual time-reversal preserving case, the magnetic dipole probes magnetic fluctuations at a wavevector $q \sim 1/z_{\text{NV}}$. We find that relaxation rates $1/T_1^{\pm \hat{z}}$ are sensitive probes of $\text{Re} \left[ \sigma_T \mp i \sigma_H \right] (q \sim 1/z_{\text{NV}}, \omega_{\text{NV}})$, where $\sigma_T$ is the optical conductivity associated with current fluctuations orthogonal to $\vs{q}$.

In Sec.~\ref{sec:qhviscosity}, we highlight how the two rates $1/T^{\pm \hat{z}}_1$ are sharply different from each other when the sample being sensed is a Hall system. The difference becomes more acute in the quantum limit with one of the two relaxation rates approaching zero. In other words, when an NV center dipole is anti-aligned with the applied magnetic field engendering the quantum Hall system, the NV center's relaxation rate is almost exclusively determined by vacuum fluctuations present even in the absence of the Hall system. We then turn our attention to making a connection with the imaginary part of the Hall conductivity at $\mathcal{O} \left( q \ell \right)^2$, where $\ell$ is the magnetic length, and the Hall viscosity. The Hall viscosity is a novel term in the viscosity tensor of a two dimensional fluid breaking time reversal symmetry~\cite{avron1995viscosity,PhysRevB.79.045308,bradlyn2012kubo}. It is quantized in a quantum Hall system and related to the Wen-Zee shift~\cite{PhysRevB.79.045308,PhysRevB.84.085316,PhysRevLett.69.953}. It has been proposed that the Hall viscosity can be measured using spatially varying fields as it appears at $\mathcal{O} \left( q \ell \right)^2$ in the \emph{real} part of the Hall conductivity, alongside another, non-topological contribution~\cite{HoyosSon}. Here we compute the imaginary part of the Hall conductivity for a non-interacting integer quantum Hall system and show that the above two contributions appear as coefficients of different poles (at $\omega = \omega_c, 2\omega_c$), and thus the Hall viscosity may be more directly inferred from the \emph{imaginary} part of the Hall conductivity, which the relaxation dynamics of the probe have access to. 

In Sec.~\ref{sec:SCs}, we consider NV relaxometry on TRSB superconductors. In particular, we study the temperature-dependent relaxation rates $\Gamma_{\pm\hat{z}}$ of the NV center in the presence of both conventional $d$-wave superconductors and chiral superconductors (using results from Refs.~\cite{lutchynchiralpwave}). We find that the average of these two rates $[\Gamma_{+\hat{z}} + \Gamma_{-\hat{z}}]/2$ exhibits a prominent Hebel-Slichter like enhancement below the superconducting transition temperature $T_c$. This is perhaps surprising given that usual NMR spectroscopy misses such a peak in $d$-wave superconductors due to the sign-changing nature of the superconducting order parameter. In the case of NV relaxometry, such a peak is present precisely because the relaxation rate is sensitive to quasiparticle scattering at small wavevectors. The difference in the two relaxation rates, $\Delta \Gamma \equiv \Gamma_{\hat{z}} - \Gamma_{-\hat{z}}$ on the other hand is only non-zero in the chiral case below $T_c$---it is directly proportional to the angular momentum of the Cooper pairs. This differential rate also inherits the Hebel-Slichter like enhacement below $T_c$. We illustrate the computation by providing experimentally expected relaxation rates in the case of topological superconductivity in stacked twisted BSCCO flakes, which have been recently investigated both theoretically and experimentally. We summarize our findings in Sec.~\ref{sec:conclusions}.

\section{Chirality of magnetic fluctuations above a time-reversal symmetry breaking material}
\label{sec:dichorism}

A standard probe of measuring time-reversal symmetry breaking in materials is the Polar Kerr (and related Faraday) effect~\cite{kapitulnik2009polar}. When linearly polarized light is normally incident on a material which exhibits magnetization or other forms of time-reversal symmetry breaking (TRSB) phenomena, the polarization of the reflected and transmitted light is generally offset by an angle that depends on the strength of TRSB in the material. One way to rationalize this is that the phase acquired upon reflection of circularly polarized light from a TRSB material is generically different for right and left circularly polarized components. This phase lag difference, which probes the magnetization of the material, or off-diagonal components of the dielectric and conductivity tensors is the cause of the rotation of the polarization~\cite{shinagawa2000faraday}. 

Likewise, the amplitude of the reflection and transmission coefficients are also generically different for the different circular polarizations of light. In particular, in ambient conditions, even in the absence of a probe light pulse, we can thus expect the amplitude of fluctuations of chiral components of the magnetic field to have different spectral properties, which could be detected by a suitable magnetic probe. Let us consider, for specificity, the example of an NV center which serves as an effective qubit with a magnetic moment $\mu_{\text{NV}}$ oriented perpendicular to the surface of a TRSB material. The NV center behaves as an effective spin$-1$ defect that has a static splitting $\omega_{\text{NV}} \approx 2\pi \cdot 3 \text{GHz}$ between $\ket{m_s = 0}$ and $\ket{m_s = \pm 1}$ levels in the absence of any external field (see Ref.~\cite{rovny2024new} and references therein). For the purposes of sensing magnetic fields which lead to transitions with $\Delta m_s = \pm 1$, we can assume that the NV center is initialized either in the state $\ket{m_s = +1}$ or in $\ket{m_s = -1}$, with the dipole oriented along the $\hat{z}$ axis, and obtain the $1/T_1$ relaxation rate by considering the decay to the $\ket{m_s = 0}$ state (using fluorescence measurements). The relevant levels, the initialized state $\ket{m_s = \pm 1}$, and the state $\ket{m_s = 0}$ can be described by a two level system Hamiltonian  

\begin{align}
H^{\pm \hat{z}}_{\text{NV}} &= \pm \frac{\hbar \omega_{\text{NV}}}{2} \sigma_z - \hbar \gamma_{\text{NV}} \vs{b} (t) \cdot \vs{\sigma} \nonumber \\
&= \pm \frac{\hbar \omega_{\text{NV}}}{2} \sigma_z - \hbar \gamma_{\text{NV}} \left(b_+ (t) \sigma_- + b_- (t) \sigma_+ + b_z (t) \sigma_z \right), 
\end{align}

where $b_\pm = \frac{b_x (t) \pm i b_y (t)}{2}$ are the chiral components of the external magnetic field at the NV center, $\sigma_\pm = \sigma_x \pm i \sigma_y$ are dimensionaless Pauli matrices, and the $\pm \hat{z}$ signs effectively impose the choice of the initial state $\ket{m_s = \pm 1}$---the state $\ket{m_s = 0}$ is associated with $\sigma_z = \mp 1$, that is, always the lower energy state,  and relaxation into the $\ket{m_s = 0}$ state from $\ket{m_s = \pm 1}$ is governed by either $b_+ \sigma_-$ (+; emitting a photon with $+1$ unit of angular momentum in the $\hat{z}$ direction), or $b_- \sigma_+$ (-; emitting a photon with $-1$ unit of angular momentum in the $\hat{z}$ direction) term accordingly. Here $\gamma_{\text{NV}}$ is the gyromagnetic ratio appropriate for the NV center. For better readability, we subsequently suppress factors of $\hbar$. We can compute the relaxation rate $1/T_1 = \Gamma_{\text{abs}} + \Gamma_{\text{ems}}$ of the NV center as the sum of the absorption and emission rates which in turn are given by Fermi's Golden rule 

\begin{widetext}
\begin{align}
\Gamma^{\pm \hat{z}}_{\text{ems}} &= 2 \pi \gamma_{\text{NV}}^2 \sum_{n,m} \rho_n \abs{\matrixel{m}{b_\pm}{n}}^2 \delta(\omega_{\text{NV}} - \omega_{mn}) \equiv \gamma_{\text{NV}}^2 |b_\pm (\omega_{\text{NV}})|^2 \nonumber \\
&= \frac{\pi}{2}  \gamma_{\text{NV}}^2  \sum_{n,m} \rho_n \left[ \abs{\matrixel{m}{b_x}{n}}^2 + \abs{\matrixel{m}{b_y}{n}}^2 \pm 2 \text{Im} \left( \matrixel{n}{b_y}{m}\matrixel{m}{b_x}{n} \right) \right] \delta(\omega_{\text{NV}} - \omega_{mn}), \nonumber \\
\Gamma^{\pm \hat{z}}_{\text{abs}} &= 2 \pi  \gamma_{\text{NV}}^2  \sum_{n,m} \rho_n \abs{\matrixel{m}{b_\mp}{n}}^2 \delta(\omega_{\text{NV}} + \omega_{mn}) \equiv \gamma_{\text{NV}}^2 |b_\mp (-\omega_{\text{NV}})|^2 \nonumber \\
&= \frac{\pi}{2}  \gamma_{\text{NV}}^2  \sum_{n,m} \rho_n \left[ \abs{\matrixel{m}{b_x}{n}}^2 + \abs{\matrixel{m}{b_y}{n}}^2 \mp 2 \text{Im} \left( \matrixel{n}{b_y}{m}\matrixel{m}{b_x}{n} \right) \right] \delta(\omega_{\text{NV}} + \omega_{mn}). \nonumber \\
\label{eq:chiralrelaxation}
\end{align}
\end{widetext}

Here $\ket{n},\ket{m}$ denote many-body eigenstates of the material, with $\hbar \omega_n, \hbar \omega_m$ their respective energies, $\omega_{mn} = \omega_m - \omega_n$ is the difference of energies modulo $\hbar$, and $\rho_n$ is the probability of the system being in state $\ket{n}$. For convenience, we set $\hbar = 1$ in what follows. Note that we make no mention of the precise kernel that connects the magnetic field to the currents/spins in the material for now, except noting that under time-reversal, these fields will flip sign, as would the current or magnetic moments generating them. It is also important to note the sign difference in the two expressions---while $\Gamma^{\hat{z}}_{\text{ems}}$ corresponds to the spectral weight of fluctuations of $b_+$ at frequency $\omega = \omega_{\text{NV}}$, $\Gamma^{\hat{z}}_{\text{abs}}$ relates to the spectral weight at $\omega = - \omega_{\text{NV}}$ of fluctuations of $b_-$. 

As seen in Eq.~(\ref{eq:chiralrelaxation}), the relaxation rates contain two types of terms, ones which involve magnetic fields in the same direction $\sim b_x^2, b_y^2$, and another set of terms comprising both the $x$ and $y$ components of the field. The latter amount to zero in a time-reversal preserving system---we can see this by noting that the matrix elements of the magnetic field for time reversed state transform as~\cite{sakurai1967advanced}

\be
\matrixel{n}{b_x}{m} \overset{\mathcal{T}}{\rightarrow} \matrixel{\tilde{n}}{b_x}{\tilde{m}} = \matrixel{n}{\mathcal{T} b_x \mathcal{T}^{-1}}{m}^* = - \matrixel{n}{b_x}{m}^*
\ee

Thus, for time-reversed states $\ket{\tilde{n}}, \ket{\tilde{m}}$ the second term $\sim b_x b_y$ changes sign as a result of complex conjugation. If the system has time-reversal symmetry, these states have the same energy as $\ket{n}, \ket{m}$ (possibly being identical to the original states), and these cross terms thus vanish. However, for a system that breaks TRSB, these terms will not in general vanish. One simple way of seeing this is to note that for a Hall system, these terms would be connected to $j_x-j_y$ current-current correlators which are linked to the presence of a finite Hall conductivity. 

By itself, the presence of these TRSB terms does not provide a clear indication of TRSB in the material, because they can in principle get swamped by the terms causing NV center relaxation even in the absence of TRSB. Moreover, while it may appear tempting to define a conclusive TRSB sensing metric by subtracting the absorption rate from the emission rate because of the sign difference in the TRSB terms, we note that, assuming the material and its ambient electromagnetic environment are in thermal equilibrium, these two rates are in fact trivially related by detailed balance~\cite{pitaevskii2016bose}. We can see this  by interchanging $m \leftrightarrow n$ in $\Gamma^{\hat{z}}_{\text{abs}}$, to recover the result that $\Gamma^{\hat{z}}_{\text{abs}} = e^{\beta \omega_{\text{NV}}} \Gamma^{\hat{z}}_{\text{ems}}$, as expected. Here we used the fact that $\rho_m = \rho_n e^{\beta \omega_{\text{NV}}}$ by virtue of the delta function and the fact that $b_x, b_y$ are Hermitian operators. In particular, $\Gamma^{\hat{z}}_{\text{abs}} \propto |b_- (-\omega_{\text{NV}})|^2$ while $\Gamma^{\hat{z}}_{\text{ems}} \propto |b_+ (\omega_{\text{NV}})|^2$.  Thus, for a system with TRSB, while the spectral functions for different chiralities of magnetic field fluctations can be different at the $\emph{same frequency}$, they are nevertheless related by detailed balance at opposite ($\omega \rightarrow - \omega$) frequencies. 

To directly probe TRSB, thus, we envision measuring the \emph{difference} of relaxation rates from the $\ket{m_s = +1}$ state (given by $\Gamma^{\hat{z}})$ and $\ket{m_s = -1}$ state (given by $\Gamma^{-\hat{z}}$. Note that for relaxation from $\ket{m_s = -1}$, $\Gamma^{-\hat{z}}_{\text{abs}} \propto |b_+ (-\omega_{\text{NV}})|^2$ while $\Gamma^{-\hat{z}}_{\text{ems}} \propto |b_- (\omega_{\text{NV}})|^2$. The difference of these two relaxation rates can only be non-zero provided the system breaks TRSB. (More generally, say for a spin-$1/2$ point defect, one may need to have two such defects, with one whose dipole points towards the system, and another whose dipole points away from the system.) Combined with the fact that the nanoscale probes are effectively point defects whose distance from the surface of the material can be controlled (for instance, by appropriate implantation depth in diamond vis-a-vis the material surface, or by mounting the probe on a cantilever), such probes can give information about the peculiar local (wave-vector dependent) scaling of correlations associated with TRSB in the material. As we show in following sections, such a diagnostic is directly sensitive to the \emph{imgainary} or dissipative part of the Hall conductivity at wave-vector $q \sim 1/z_{\text{NV}}$, where $z_{\text{NV}}$ is the distance of the probe from the material surface.

To complete this discussion, it is useful to define symmetrized and anti-symmetrized spectral functions and related susceptibilities. Specifically, we introduce 
\begin{widetext}
\begin{align}
N_{xx (yy)} (\omega > 0) &= \pi \sum_{n,m} \rho_n \abs{\matrixel{m}{b_{x(y)}}{n}}^2 \left[ \delta(\omega - \omega_{mn}) + \delta (\omega + \omega_{mn} ) \right], \nonumber \\
N_{yx} (\omega > 0) &= \pi \sum_{n,m} \rho_n \matrixel{n}{b_y}{m} \matrixel{m}{b_x}{n} \left[ \delta(\omega - \omega_{mn}) - \delta(\omega + \omega_{mn}) \right], \nonumber \\
A_{xx} (\omega > 0) &= \pi \sum_{n,m} \rho_n  \abs{\matrixel{m}{b_x}{n}}^2 \left[ \delta(\omega - \omega_{mn}) - \delta(\omega + \omega_{mn}) \right], \nonumber \\
A_{yx} (\omega > 0) &= \pi \sum_{n,m} \rho_n \matrixel{n}{b_y}{m} \matrixel{m}{b_x}{n} \left[ \delta(\omega - \omega_{mn}) + \delta(\omega + \omega_{mn}) \right]. \nonumber
\end{align}

These spectral functions appear in the relaxation rate as
\begin{align}
\Gamma_{\pm \hat{z}} \equiv \frac{1}{T^{\pm \hat{z}}_1} &=  \gamma_{\text{NV}}^2 \left[ \frac{1}{2} N_{xx} (\omega_{\text{NV}}) + \frac{1}{2} N_{yy} (\omega_{\text{NV}}) \pm \text{Im}[ N_{yx} (\omega_{\text{NV}}) ] \right], \nonumber \\
\frac{1}{T^{\pm\hat{z}}_1} &=  \gamma_{\text{NV}}^2 \coth \left( \frac{\beta \omega_{\text{NV}}}{2} \right) \left[ A_{xx} (\omega_{\text{NV}} ) \pm  \text{Im}[ A_{yx} (\omega_{\text{NV}}) ] \right], \nonumber \\
\frac{1}{T^{\hat{z}}_1} &- \frac{1}{T^{-\hat{z}}_1} = 2  \gamma_{\text{NV}}^2 \coth \frac{\beta \omega_{\text{NV}}}{2} \text{Im} \left[ A_{yx} (\omega_{\text{NV}}) \right],
\label{eq:T1}
\end{align}
\end{widetext}

where we have assumed that the system respects rotational invariance and thus $A_{xx} (\omega) = A_{yy} (\omega)$. The $\text{coth}$ factors obtained reflect the fluctuation-dissipation theorem. Note that $N_{xy}$ is only non-zero due to time-reversal symmetry breaking. Concurrently, we define the susceptibility

\begin{align}
\chi_{b_{y(x)}, b_x} &= \mathcal{F} \left[ - i \theta(t) \avg{ \left[ b_{y(x)} (t), b_x (0) \right]} \right] \nonumber \\
&= - \sum_{n,m} \rho_n \bigg[ \frac{\matrixel{n}{b_{y(x)}}{m}\matrixel{m}{b_x}{n}}{\omega- \omega_{\text{mn}} + i0^+} \nonumber \\
& \hspace{1in} - \frac{\matrixel{n}{b_x}{m}\matrixel{m}{b_{y(x)}}{n}}{\omega + \omega_{\text{mn}} + i0^+}  \bigg], 
\label{eq:chi}
\end{align}

which will be easier to compute going forward and which are related to the anti-symmetric correlators $A_{y(x) x} (\omega)$ via the relations---

\begin{align}
    \text{Im} \left[ \chi_{b_x,b_x} (\omega) \right] &= A_{xx} (\omega) = \tanh \left( \frac{\beta \omega}{2} \right) N_{xx}, \nonumber \\
    \text{Re} \left[ \chi_{b_y, b_x} (\omega) \right] &= - \text{Im}\left[ A_{yx} (\omega) \right] = - \tanh \left( \frac{\beta \omega}{2} \right) \text{Im} \left[ N_{yx} \right].
\end{align}

Note that in the result above, one may be concerned that there are contributions to $\text{Re} \left[ \chi_{b_y, b_x} \right]$ from terms proportional to $\text{Re} \left[ \matrixel{n}{b_y}{m} \matrixel{m}{b_x}{n} \right] \mathcal{P} \left( \frac{1}{\omega \pm \omega_{mn}} \right)$. There are two reasons for excluding such contributions---i) these should be much smaller than the terms proportional to $\delta (\omega - \omega_{mn})$ corresponding to $\text{Im} \left[ A_{yx} \right]$, and ii) in the case of the system possessing rotational symmetry, they vanish exactly since $\chi_{b_y, b_x} = - \chi_{b_x, b_y}$ while these terms are symmetric under the exchange $b_x \leftrightarrow b_y$. The latter can be easily intuited by noting that $\chi_{b_y,b_x}$ corresponds to the magnetic field $b_y$ in response to a magnetic moment placed at the site of the probe with moment in the $x$ direction, and considering a $\pi/2$ rotation. 

We also note that an analogous measurement of $\text{Im} [ N_{yz} ]$ or $\text{Im} [ N_{xz} ]$ is possible by orienting the dipole moment accordingly but here we focus for simplicity exclusively on $\text{Im} [ N_{xy} ]$.

\section{Measurement of chirality in conducting materials}
\label{sec:sigmaxy}

We now evaluate $\text{Im} \left[ N_{yx} (\omega) \right]$ by solving the problem of reflection of light from a two-dimensional metallic surface sandwiched between two dielectrics with the same dielectric constant $\epsilon$; see Fig.~\ref{fig:trsbnoise} for the setup. We follow the methods of Ref.~\cite{agarwal2017magnetic}, but allow for the two dimensional material to exhibit a Hall response owing to TRSB. To evaluate $\text{Im} \left[ N_{yx} (\omega) \right]$, we first compute the susceptibility $\chi_{b_y, b_x} (\vs{r}) = B^{\text{total}}_y (\vs{r}) / M_x (\vs{r})$ by computing the electromagnetic field set up by a local magnetic dipole at site of the probe $\vs{r} \equiv (z_{\text{NV}}, \vs{\rho}_\text{NV}) $ in the presence of the material which breaks TRSB. 

To simplify this task, we use the translation symmetry along the surface of the material and calculate the magnetic field profile due to a sheet of magnetization, with an in-plane magnetic moment oscillating with a fixed in-plane wavevector $\vs{Q}$. To obtain the for a point dipole moment, we add the responses from each such magnetization sheet with the appropriate inverse Fourier factor. Specifically, we solve for the magnetic field in the presence of a sheet of magnetization  $\vs{M} (z, \vs{\rho}) = m_0 (\hat{z} \times \hat{Q}) \delta(z - z_{NV} ) e^{i \vs{Q} \cdot \vs{\rho} }$ and, separately, in the presence of a sheet of magnetization with $\vs{M} (z, \vs{\rho}) = m_0 \hat{Q} \delta(z - z_{NV} ) e^{i \vs{Q} \cdot \vs{\rho} }$. Here $\vs{\rho}$ is the in-plane component of the position vector $(z, \vs{\rho})$ and $\vs{Q}$ is an arbitrary in-plane wavevector which we will ultimately integrate over to obtain the response due to a finite magnetic moment at the site of the NV center, whose coordinates are $z = z_{\text{NV}}, \vs{\rho} = \vs{\rho}_{\text{NV}}$. 

In particular, the total magnetic field due to the point dipole moment can then be found by integrating the magnetic field found (in the $y$-direction) in these two separate calculations with the Fourier factor $e^{-i \vs{Q} \cdot \vs{\rho}_{\text{NV}}}$ and using appropriate projection to $\hat{x}$, $\hat{y}$ axes. The decomposition into moments parallel to $(\hat{z} \times \hat{Q})$ and $\hat{Q}$ is done because magnetization sheets in these directions yield exclusively $p-$ (magnetic field is in-plane) and $s-$ polarized waves (electric field is in-plane), respectively. In the case of systems that do not break TRSB, it is known that the reflected waves continue to retain their polarization; see Ref.~\cite{agarwal2017magnetic} for the related calculation in the absence of TRSB. However, in this case, as we show, the presence of a Hall conductivity generates waves of opposite polarization as well. 

\subsubsection{s-polarized incident waves solution}

The electric field is in-plane and the magnetic field can be found using Faraday's Law, $\vs{B} = \curl{E}/(i \omega)$. The general form of the electromganetic fields is given by

\begin{align}
\vs{E}_{in} &= E_0 \left( \hat{z} \times \hat{Q} \right) e^{i \vs{Q} \cdot \vs{\rho} - i q_z z}, \nonumber \\
\vs{E}_{r} &= E_0 \left[ r_{ss} \left(\hat{z} \times \hat{Q} \right) + r_{sp}( -\bar{q}_z \hat{Q} + \bar{Q} \hat{z} ) \right] e^{i \vs{Q} \cdot \vs{\rho} + i q_z z}, \nonumber \\
\vs{E}_{t} &= E_0 \left[t_{sp} (\bar{q}_z \hat{Q} + \bar{Q} \hat{z} )+t_{ss} \left( \hat{z} \times \hat{Q} \right) \right]e^{i \vs{Q} \cdot \vs{\rho} - i q_z z}. \nonumber \\
\vs{B}_{in} &= \frac{E_0}{ \bar{c}} \left( \bar{q}_z \hat{Q} + \bar{Q} \hat{z} \right) e^{i \vs{Q} \cdot \vs{\rho} - i q_z z}, \nonumber \\
\vs{B}_{r} &= \frac{E_0}{\bar{c}} \left[ r_{ss} \left( - \bar{q} \hat{Q} +  \bar{Q} \hat{z} \right)  - r_{sp} (\hat{z} \times \hat{Q} ) \right] e^{i \vs{Q} \cdot \vs{\rho} + i q_z z}, \nonumber \\
\vs{B}_{t} &= \frac{E_0}{\bar{c}} \left[ t_{ss}\left( \bar{q}_z \hat{Q} + \bar{Q} \hat{z} \right) - t_{sp} (\hat{z}\times \hat{Q}) \right] e^{i \vs{Q} \cdot \vs{\rho} - i q_z z}. \nonumber \\
\end{align}

where $q_z = \sqrt{\epsilon \frac{\omega^2}{c^2} - Q^2}$, $Q = \norm{\vs{Q}}$ and we have defined $\bar{c} = c/\sqrt{\epsilon}$ as the speed of light in the dielectric, and dimensionless wavevectors $\bar{q}_z = q_z \bar{c}/\omega, \bar{Q} = Q \bar{c}/\omega$. Thus, $\bar{q}_z^2 + \bar{Q}^2 = 1$. The coefficients $r_{ss}, r_{sp}, t_{ss}, t_{sp}$ are reflection and transmission coefficients, respectively, for the case of when $s$-polarized light impinges on the sample. The subscript $sp$ and $ss$ denote generation of $p$-polarized and $s-$ polarized light, respectively, from incident $s$-polarized light. Note that fields can be evanascent if $q_z$ is imaginary, but are propagating in plane (given in-plane translational symmetry). 

To solve for the reflection and transmission coefficients, we apply the following boundary conditions: the fields $B_z$ and $\epsilon \vs{E}_\parallel$ are continuous across the interface, while the parallel component of $\vs{B}$ and $E_z$ are discontinuous due to the free charge and current generated at the interface, that is, $\mu_0 \vs{J} = \hat{z} \times \left( \vs{B}_\parallel (0^+) - \vs{B}_\parallel (0^-) \right)$, and $\rho / \epsilon_0 = \epsilon E_z (0^+) - \epsilon E_z (0^-)$. The material imposes a constraint between the charge accumulated and the longitudinal current via the continuity relation, while the current itself is given by the conductivity tensor of the material; specifically, $\vs{J} = \sigma_L \vs{E}_{\parallel, \hat{Q}} + \sigma_T \vs{E}_{\parallel, (\hat{z} \times \hat{Q})} - \sigma_H (\hat{z} \times \vs{E}_\parallel)$. Note in particular the sign of $\sigma_H$ is determined by the definition $J_x = \sigma_H E_y$. Here we make a distinction between current response that is parallel or perpendicular to the modulation wave-vector $\vs{Q}$ denoting these by $\sigma_L$ and $\sigma_T$ respectively. 

Putting the above together, and defining dimensionless conductivities $\bar{\sigma}_{H/L/T} = \sigma_{H/L/T} / \left( \bar{c} \epsilon_0 \epsilon \right)$, we obtain the following consistency conditions on the reflection and trasmission coefficients: 

\begin{align}
1 + r_{ss} &= t_{ss}, \; \; r_{sp} = - t_{sp}, \nonumber \\
t_{sp} \bar{\sigma}_L \bar{q}_z + t_{ss} \bar{\sigma}_H &=  r_{sp} - t_{sp}, \nonumber \\
t_{ss} \left(\bar{\sigma}_T + \bar{q}_z \right) &= \bar{q}_z (1 - r_{ss}) + \bar{\sigma}_H \bar{q}_z t_{sp}. 
\end{align}

Solving for $r_{sp}$ and $r_{ss}$, we find

\begin{align}
1 + r_{ss} &= \frac{1}{1 + \frac{\bar{\sigma}_T}{2 \bar{q}_z} + \frac{\bar{\sigma}_H}{2 \bar{q}_z} \cdot \frac{\bar{\sigma}_H}{2 + \bar{q}_z \bar{\sigma}_L}} \approx 1 - \frac{\bar{\sigma}_T}{2 \bar{q}_z} \nonumber \\
r_{sp} &= \frac{\bar{\sigma}_H ( 1 + r_{ss} )}{2 + \bar{q}_z \bar{\sigma}_L } \approx \frac{\bar{\sigma}_H}{2 + \bar{q}_z \bar{\sigma}_L }
\end{align}

Here we note that for $\sigma \sim e^2/h$, $\bar{\sigma}$ is generically a small number $\sim 0.1$, and wavevectors $q_z \sim 1/[10 \text{nm}]$ and frequency of interest $\omega = \omega_{\text{NV}} \sim \text{GHz}$, $\bar{q}_z \sim 10^7$. This justifies the approximations made above. 

\subsubsection{p-polarized incident waves solution}

The magnetic field is now in-plane and the electric field is found using the relation $\vs{E} = \frac{\boldsymbol{\nabla} \times \vs{B} c^2}{-i \omega \epsilon}$. The fields are

\begin{align}
\vs{B}_{in} &= B_0 \left( \hat{z} \times \hat{Q} \right) e^{i \vs{Q} \cdot \vs{\rho} - i q_z z}, \nonumber \\
\vs{B}_{r} &= B_0 \left[ r_{pp} \left(\hat{z} \times \hat{Q} \right) + r_{ps} \left( -\bar{q}_z \hat{Q} + \bar{Q} \hat{z} \right) \right] e^{i \vs{Q} \cdot \vs{\rho} + i q_z z}, \nonumber \\
\vs{B}_{t} &= B_0 \left[ t_{pp} \left( \hat{z} \times \hat{Q} \right) + t_{ps}  \left( \bar{q}_z \hat{Q} + \bar{Q} \hat{z} \right) \right] e^{i \vs{Q} \cdot \vs{\rho} - i q_z z}.\nonumber \\
\vs{E}_{in} &= -\bar{c}B_0\left( \bar{q}_z \hat{Q} + \bar{Q} \hat{z} \right) e^{i \vs{Q} \cdot \vs{\rho} - i q_z z}, \nonumber \\
\vs{E}_{r} &= \bar{c} B_0 \left[ r_{pp} \left(\bar{q}_z \hat{Q} - \bar{Q} \hat{z} \right) + r_{ps} (\hat{z} \times \hat{Q} )\right]  e^{i \vs{Q} \cdot \vs{\rho} + i q_z z}, \nonumber \\
\vs{E}_{t} &=  \bar{c} B_0 \left[t_{ps} (\hat{z} \times \hat{Q} ) - t_{pp} \left(\bar{q}_z \hat{Q} + \bar{Q} \hat{z} \right) \right] e^{i \vs{Q} \cdot \vs{\rho} - i q_z z}. \nonumber \\
\end{align}

The boundary conditions yield the consistency conditions 
\begin{align}
1- r_{pp} &= t_{pp}, \; \; r_{ps} = t_{ps}, \nonumber \\
\bar{\sigma}_T t_{ps} + \bar{q}_z \bar{\sigma}_H  t_{pp} &= - \bar{q}_z (r_{ps} + t_{ps}) \nonumber \\
\bar{\sigma}_H t_{ps} - \bar{q}_z \bar{\sigma}_L t_{pp} &= -(1+r_{pp}) + t_{pp}
\end{align}

Solving this gives for $r_{pp}, r_{ps}$ yields

\begin{align}
r_{pp} &= \frac{\bar{q}_z \bar{\sigma}_L + \frac{\bar{q}_z \bar{\sigma}^2_H}{\bar{\sigma}_T + 2 \bar{q}_z} }{2 + \bar{q}_z \bar{\sigma}_L + \frac{\bar{q}_z \bar{\sigma}^2_H}{\bar{\sigma}_T + 2 \bar{q}_z} } \approx \frac{\bar{q}_z \bar{\sigma}_L}{2 + \bar{q}_z \bar{\sigma}_L} \nonumber \\
r_{ps} &= \frac{- \bar{\sigma}_H}{\left( 1 + \frac{\bar{\sigma}_T}{2\bar{q}_z} \right) ( 2 + \bar{q}_z \bar{\sigma}_L) + \frac{1}{2} \bar{\sigma}_H^2} \approx - \frac{\bar{\sigma}_H}{2 + \bar{q}_z \bar{\sigma}_L}
\end{align}

where we again neglect higher order contributions in $\bar{\sigma}/\bar{q}$ as justified above.

\subsubsection{Computation of $\text{Im} \left[ N_{xy} \right]$}

To obtain the susceptibility $\chi_{yx}$, we now combine the results from the preceding computations as follows. We would like to compute the magnetic field generated as a consequence of a dipole of strength $m_0$ placed as the site of the probe in the $x-$ direction, and compute the $y-$ component of the reflected magnetic field $B_{r,y}$ to compute the connected (non-vacuum) part of $\chi_{yx} = B_{r,y}/m_0$. Here, we first compute the magnetic field due to a magnetic sheet $\vs{M} = m_0 \hat{x} \delta (z - z_{\text{NV}}) e^{i \vs{Q} \cdot \vs{\rho}}$ instead of a point dipole. To obtain the result for a point dipole, we can integrate over $\vs{Q}$ with the appropriate Fourier factor. 

This magnetization sheet generates a p-polarized field with strength $B_0 = i \mu_0 m_0 \epsilon \frac{\omega^2 /c^2}{2 q_z} \hat{x}.(\hat{z} \times \hat{Q}) e^{i q_z z_{\text{NV}}}$, and an s-polarized component with $E_0 = - \mu_0 \omega m_0 \frac{Q}{2 q_z} \hat{x} \cdot \hat{Q} e^{i q_z z_{\text{NV}}}$. Note that the reflected field consists of in-plane components parallel and perpendicular to $\vs{Q}$. It is easy to see that only components of the reflected field coming from $r_{sp}$, $r_{ps}$ survive angular averaging over $\hat{Q}$ and contribute to $\chi_{yx}$. (For $\chi_{xx}, \chi_{yy}$, it is the components of the reflected field proportional to $r_{ss}, r_{pp}$ that contribute.) Additionally, we note that the wave-vector $q_z \approx i Q$. This is because at small probe-material distances, the phase space of large $Q \sim 1/r_{\text{probe}}$ contributes most to the susceptibility, while the frequency $\omega$ is generically small, such that $\epsilon \omega^2 / c^2 \ll Q^2$. Thus, the response largely comes from evanescent waves for which $q_z$ is purely imaginary. Combining this fact with the results of the preceding subsections, we find

\begin{align}
\chi_{yx} &= i \int \frac{dQ}{8 \pi}  e^{-2 Q z_{NV}} \frac{ \mu_0 \omega Q}{\bar{c}} (r_{ps} - r_{sp}), \nonumber \\
\text{Im}\left[ N_{yx} \right] &= \coth{\frac{\beta \omega}{2}} \int \frac{d Q}{8 \pi} e^{-2 Q z_{NV}} \frac{\mu_0 \omega Q}{\bar{c}} \text{Im} \left[ r_{ps} - r_{sp} \right], \nonumber \\
&\approx -\frac{\mu^2_0 \omega \coth{\frac{\beta \omega}{2}}}{32 \pi z^2_{NV}} \int dx \; x e^{-x} \text{Im} \left[ \frac{\sigma_H}{\epsilon_{\text{RPA}}}\right] \left(\frac{x}{2z_{\text{NV}}} , \omega \right) , \nonumber \\
\label{eq:sigmaxymeasure}
\end{align}

where $\epsilon_{\text{RPA}} = 1 + i \frac{Q \sigma_L}{2 \epsilon \epsilon_0 \omega}$ is the usual RPA screening factor at finite wavevector $Q$ and frequency $\omega$ in two dimensions. The integral is evaluated over the range $ x = 0$ to $x = \infty$.  

It is important to note that the relaxation dynamics of the probe comes from the \emph{imaginary} part of $\sigma_H$ since it is this piece that is dissipative. Furthermore, as suggested before, much of the response comes from the wavevector $Q \approx 1/2 z_{\text{NV}}$---lower wavevectors have a smaller density of states, suppressed by an additional factor of $Q$, while higher wavevectors are averaged out (as reflected in the exponential factor $e^{-x}$ in Eq.~(\ref{eq:sigmaxymeasure}). 

\subsubsection{Computation of $\text{Re} \left[ N_{xx (yy)} \right]$}

To obtain the susceptibility $\chi_{xx}$, we compute the reflected magnetic field $\vs{B}_r$ in the $x-$ direction due to a magnetic dipole of strength $m_0$ placed as the site of the probe in the $x-$ direction. The result for $\chi_{yy}$ follows by rotational symmetry. We follow the same steps as above to arrive at the following result

\begin{align}
    \chi_{xx} &= \chi_{yy} = \int \frac{dQ}{8 \pi}  e^{-2 Q z_{NV}} \mu_0 Q^2 (r_{ss} + \frac{\omega^2}{\bar{c}^2 Q^2} r_{pp}), \nonumber \\
    \chi_{xx} &= \chi_{yy} \approx i \frac{\mu^2_0 \omega}{64 \pi z^2_{\text{NV}}} \int dx \; x e^{-x} \left( \sigma_T + \frac{\sigma_L}{\epsilon_{\text{RPA}}} \right) \nonumber \\
    N_{xx} &= N_{yy} \approx \frac{\mu^2_0 \omega \coth {\beta \omega/2}}{64 \pi z^2_{\text{NV}}} \int dx \; x e^{-x} \text{Re} \left[ \sigma_T + \frac{\sigma_L}{\epsilon_{\text{RPA}}} \right]
\end{align}

where we note that the $\sigma_T, \sigma_L, \epsilon_{\text{RPA}}$ are evaluated at $q = x/2z_{\text{NV}}, \omega = \omega_{\text{NV}}$ and we have assumed that $q \approx 1/2z_{\text{NV}} \gg \omega_{NV}/c$. The integral is evaluated over the range $ x = 0$ to $x = \infty$.  

\subsubsection{Combined result for the $1/T_1$ relaxation rate}

We now combine the above results to arrive at the final result for the relaxation rate

\begin{align}
\Gamma_{\pm \hat{z}} \approx \frac{\mu^2_0 \omega \coth {\beta \omega/2}}{64 \pi z^2_{\text{NV}}} \int dx \; x e^{-x} \text{Re} \left[\sigma_T + \frac{\sigma_L}{\epsilon_{\text{RPA}}} \pm i \frac{2\sigma_H}{\epsilon_{\text{RPA}}} \right]
\end{align}

We note that in general, the presence of the factor $\epsilon_{\text{RPA}}$ makes the interpretation of the noise measurement somewhat more non-trivial. This factors arises for both $\sigma_L$ and $\sigma_H$ terms but not for the $\sigma_T$ term---this is because the former involve the presence of currents with a non-zero divergence while the latter only corresponds to a divergence-free current which is thus not screened. Indeed, for $q \ell_{\text{sc}} \ll 1$, where $\ell_{\text{sc}}$ is the screening length, this term will polynomially suppress the magnetic noise originating from longitudinal and Hall current fluctuations. However, for $q \ell_{\text{sc}} \gtrsim 1$, we can approximately ignore this effect and set $\epsilon_{\text{RPA}} = 1$ in the above. 

In the next section, for simplicity we will focus on the situations where $q \ell \lesssim 1, q \ell_{\text{sc}} \gtrsim 1$. Here $\ell$ is a general length scale governing the TRSB phenomena and a scale which governs the gradient expansion of $\sigma_H$. In the pure quantum Hall case, this will refer to the magnetic length. In the case of TRSB superconductors, or paired composite fermions in a fractional quantum Hall system, this length scale will be given by $v_F \hbar/ \Delta$, where $\Delta$ is the pairing gap. In this limit, one can approximate $\epsilon_{\text{RPA}} \approx 1$, while also setting $\sigma_L \approx \sigma_T$. We then obtain the simplified result

\begin{align}
    \Gamma_{\pm \hat{z}} \approx \frac{\mu^2_0 \omega \coth {\beta \omega/2}}{32 \pi z^2_{\text{NV}}} \int dx \; x e^{-x} \text{Re} \left[\sigma_T \pm i \sigma_H \right]. 
\end{align}

Note that, regardless of the suppression of noise by screening, it always remains possible to isolate the noise proportional to the Hall conductivity by subtracting the relaxation rates of the NV centers initialized in the $\ket{m_s = \pm 1}$ states. Thus, 

\begin{align}
    \Gamma_{\hat{z}} - \Gamma_{-\hat{z}} \approx -\frac{\mu^2_0 \omega \coth {\beta \omega/2}}{16 \pi z^2_{\text{NV}}} \int dx \; x e^{-x} \text{Im} \left[\frac{\sigma_H}{\epsilon_{\text{RPA}}} \right]
\end{align}

\section{Quantum Hall and Hall viscosity measurements}
\label{sec:qhviscosity}

The discrepancy in the decay rate of the dipole probe pointing in the $+\hat{z}$ and $-\hat{z}$ directions perpendular to a material surface is most pronounced in the Hall state, particularly in the quantum Hall limit. To see this, let us first consider the Hall system in the classical limit where the Drude result applies. Here, in the limit $q \rightarrow 0$, we obtain the result

\begin{align}
    \sigma_{T} &= \sigma_0 \frac{1 - i \omega \tau}{1 + (\omega_c^2 - \omega^2) \tau^2 - 2 i \omega \tau}, \nonumber \\
    \sigma_H &= \sigma_0 \frac{-\omega_c \tau}{1 + (\omega^2_c - \omega^2) \tau^2 - 2i \omega \tau}, 
\end{align}

where $\sigma_0 = ne^2\tau/m$ is the usual zero-field Drude result at zero frequency, $\omega_c$ is the cylctron frequency, and $\tau$ is the current relaxation rate. 

From the Drude result, we obtain

\be
    \left[ \sigma_T \pm i \sigma_H \right] (q = 0, \omega) = \sigma_0 \left(\frac{1 - i (\omega \pm \omega_c) \tau}{1 + (\omega^2_c - \omega^2) \tau^2 - 2i \omega \tau} \right), 
\ee

Clearly, one has for the real part of the appropriate conductivity in the two cases at $\omega = \omega_c$, 

\be
\text{Re} \left[ \sigma_T \pm i \sigma_H \right] (\omega = \omega_c)= \sigma_0 (+\text{ve}), \frac{\sigma_0}{1 + (2 \omega_c \tau)^2} (-\text{ve}).
\ee 

In the quantum Hall limit, where $\omega_c \tau \gg 1$, one finds that the noise contribution at $q = 0$ tends to go to zero in one case, while it is finite in the other. (See also Ref.~\cite{PhysRevLett.129.046801} for an experiment that probes the optical conductivity of a quantum Hall system by confining it to an appropriate cavity observes a related effect.) Thus, a probe separated at distances much larger than the magnetic length from the material will exhibit virtually no relaxation due to the presence of a Hall bar when oriented in the $-\hat{z}$ direction (in alignement with the external field) while experiencing a significant contribution to the relxation from the presence of the material when placed in the opposite direction, at low temperatures. Another simple way of seeing this result emerge is by noting that the magnetic field component $b_\pm$ can be connected to the appropriate current matrix elements $j_\pm = j_x \pm i j_y$ in the material. These current matrix elements only serve to increase or decrease the Landau level index, respectively, at $q = 0$. In particular, $j_x + ij_y$ serves to increase the Landau level index, and its fluctuations at zero temperature are finite, while those of $j_x - ij_y$ are small due to Fermi blocking. Consequently, the relaxation of the probe placed in the $+\hat{z}$ direction, which is due to $b_+$ fluctuations, will be finite, while in the case the probe dipole is placed in the $-\hat{z}$ direction, the relevant fluctuations are that of $b_-$, which will approach zero as the temperature is lowered. 

\subsection{Finite $q$ response and Hall viscosity}

The Hall viscosity is a novel non-dissipative kind of viscous effect that appears in quantum Hall systems. In general, viscosity is a tensor associated with the stress in a fluid due to a time-dependent strain

\be
P_{ij} = \sum_{kl} \eta_{ij, kl} v_{kl}, 
\ee

where $i,j \in \{x,y\}$, $v_{kl} = \frac{1}{2} \left( \partial_k v_l + \partial_l v_k \right)$ is the symmetrized gradient of the velocity field $\vs{v}$ and $P_{ij}$ is the associated stress. In homogeneous systems, with time-reversal invariance, two scalar coefficients (sheer $\eta$ and bulk $\xi$ viscosities) completely define the viscosity tensor. However, for a two-dimensional system, TRSB allows for the existence of a third non-dissipative component known as Hall viscosity, denoted as $\eta_H$. In $d = 2$, the full viscosity tensor is given by

\begin{align}
\eta_{ij, kl} &= \xi \delta_{ij} \delta_{kl} + \eta( \delta_{ik} \delta_{jl} + \delta_{il} \delta_{jk} - \delta_{ij} \delta_{kl} ) + \nonumber \\
&\frac{\eta_H}{2} (\epsilon_{ik} \delta_{jl} + \epsilon_{il} \delta_{jk} + \epsilon_{jk} \delta_{il} + \epsilon_{jl} \delta_{ik} ),
\end{align} 

where $\delta_{jk}$ is the Kronecker delta tensor and $\epsilon_{jk}$ is the Levi-Civita tensor in two dimensions. The Hall viscosity term produces a force density $\vs{f} = \eta_H \nabla^2 \vs{v} \times \hat{z}$---it is relevant near the local maxima and minima of the velocity field in the fluid and produces a force perpendicular to the field. In particular, $\eta_{H} = \eta_{xx, xy}$, and its real part can be evaluated by computing the linear response coefficient~\cite{bradlyn2012kubo}

\be
\eta_H = \lim_{\omega \rightarrow 0} \frac{\text{Im} \chi_{P_{xx} , P_{xy}}}{\omega}.
\ee 

[$\chi_{P_{xx},P_{xy}}$ represents the Kubo response function defined analogously to Eq.~(\ref{eq:chi})] The application of strain can be viewed as equivalent to deformations of the metric. The deformation of wave-functions under such a change can be associated with a Berry curvature whose integral is quantized~\cite{avron1995viscosity}. The Hall viscosity defined in this sense is therefore a useful quantity to study; it has been shown to distinguish various candidate gapped quantum Hall states such as the Pfaffian and Anti-Pfaffian~\cite{PhysRevB.84.085316}, is proportional to the pairing angular momentum ($l = \pm 1$) in chiral p-wave superfluids~\cite{bradlyn2012kubo}, and also can be related to the global curvature when studying quantum Hall systems on curved manifolds (the Wen-Zee shift~\cite{wenzee}). Due to this, it has garnered significant recent theoretical attention, but experimentally it remains a challenge to measure.  

Hoyos and Son~\cite{HoyosSon} showed for Galilean-invariant systems that the Hall viscosity in fact appears in the wavevector ($q$-) dependent Hall conductivity $\text{Re}\left[\sigma_{H}\right]$ at order $\mathcal{O} \left( (q \ell)^2 \right)$. Here $\ell = \sqrt{\hbar c/eB}$ is the magnetic length. In general, since a time-dependent strain is associated with spatially varying velocity fields, one may imagine that a spatially-varying electric field may produce some response proportional to this viscous effect. Indeed, Hoyos and Son find

\be
\frac{\sigma_{xy} (q)}{\sigma_{xy} (0) } = 1 + C_2 \cdot (q \ell)^2 + \mathcal{O} (q^4 \ell^4), 
\ee  

where 

\be
C_2 = \frac{\eta_H}{\hbar n} - \frac{2\pi}{\nu} \frac{\ell^2}{\hbar \omega_c} B^2 \epsilon'' (B). 
\ee

With $n$ being the electron density, $\nu$ is the filling fraction and $\epsilon(B)$ is the energy density. The first quantity (the precise ratio) is quantized, while the second quantity is another response function which is not quantized but can be measured independently by applying a space-dependent magnetic field. The result can be intuited as follows. First, an electric field $E_x$ varying along the $x-$ direction with a wave-vector $q$ produces via the Hall effect a spatially varying velocity profile $v_y (x)$ in the $y-$direction; this amounts to a viscous stress $\sim q$ in the $x$ direction. The force density $\sim q^2$ then produces (via the Hall conductance coefficient) an additional current $v_y$ which is proportional to the $q^2$. This is the quantized part of the response at $q^2$. Another part of the response is due to the fact that non-zero $\partial_x v_y$ is associated with an effective magnetic field (due to the circulation of the current) $\sim q$ which produces a magnetization related to the magnetic susceptibility of the Quantum Hall fluid. This magnetization corresponds to additional current density given by the curl of this magnetization (and is thus, $\sim q^2$). The second term is usually of the order of the first term so they are not easy to separate. Measuring the viscosity from the real part of the Hall conductivity can thus be experimentally challenging because it is hard to measure the difference between two different response functions separately and accurately. 

The local probes we study are particularly useful for studying local, finite wave-vector response and it is thus pertinent to ask if the Hall viscosity could be determined from the relaxation rate of such probes. However, relaxation is by definition a process that is dissipative, which is why the probes are sensitive to the imaginary part of the Hall conductivity. We are thus led to ask if signatures of the Hall viscosity term show up in the \emph{imaginary} part of the Hall conductivity at finite wavevectors. In this work, we consider the simplest scenario---that of integer quantum Hall states in Gallilean-invariant 2DEG and in graphene. In both instances, we find the answer to the above question is in the affirmative, and that additionally, the two terms in $C_2$ are associated with different frequency poles, with the Hall viscosity term appearing along with the pole $\omega = 2 \omega_c$, while the non-local magnetization current appears at $\omega = \omega_c$. Here $\omega_c$ is the cyclotron frequency separating Landau levels. 

\subsection{Computation of the Hall viscosity for the integer Quantum Hall state in 2DEGs}

In this section, we compute the i) Hall viscosity from the appropriate Kubo response function, ii) Real and imaginary parts of frequency dependent $\sigma_T (q, \omega), \sigma_H (q, \omega)$ to $\mathcal{O} \left( q \ell \right)^2 $ and show that i) the $\pm$ combinations $\left[ \sigma_T \pm i \sigma_H \right]$ exhibit very different responses, with one of the terms vanishing at order $\mathcal{O} \left( q \ell \right)^0$, ii) confirm that $\text{Re} \left[ \sigma_H (q, \omega) \right]$ yields a term proportional to the Hall viscosity and to the magnetic subceptibility at $\mathcal{O} \left( q \ell \right)^2$ as advertised in previous works, and iii) show that the $\text{Im} \left[ \sigma_H (q, \omega) \right]$ exhibits poles at $\omega = \omega_c, 2 \omega_c$ where the Hall viscosity part appears exclusively at $\omega = 2 \omega_c$. 

We consider the case of the integer quantum Hall states in a non-interacting 2DEG, with a spatially uniform metric tensor $g_{ij}$. The Hamiltonian is given by

\be
H = \frac{1}{2m} \sum_{ij} g_{ij} \frac{\{ \Pi_i , \Pi_j \}}{2},
\ee

where $\Pi_i = p_i + \frac{e}{c} A_i$ are the canonical momenta for $i = x,y$. For simplicity, we work in the Landau gauge with $\vs{A} = x B \hat{y}$. In what follows, we set $\ell = 1, \hbar = 1, \omega_c = 1$ and restore these in the final results. 

The symmetrized stress operator $P_{ij}$ relating deformations in the $j-$th direction varying in the $i-$th direction spatially is given by $P_{ij} = \frac{2}{\sqrt{g}} \pd{H}{g_{ij}}$, where $g = \text{det}[g_{ij}]$. In what follows, we will work with flat space metric with $g_{ij} = \delta_{ij}$. The relevant stress operators for computing the Hall viscosity are 

\be
    P_{xx} = \Pi^2_x, \; \; P_{xy} = \frac{\{\Pi_x, \Pi_y\}}{2}.
\ee

The finite wave-vector current operators of interest are given by

\begin{align}
    J_y (q) &= \frac{1}{2}\left\{ j_y, e^{-iqy} \right\} = \frac{-e}{2} \left\{ \Pi_y, \left( 1 + i q \Pi_x - \frac{q^2}{2} \Pi_x^2 \right) \right\},\nonumber \\
    J_x (q) &= \frac{1}{2}\left\{ j_x, e^{-iqy} \right\} = \frac{-e}{2} \left\{ \Pi_x, \left( 1 + i q \Pi_x - \frac{q^2}{2} \Pi_x^2 \right) \right\},
\end{align}

where we used coordinate-momentum locking in Landau level states to replace $e^{-iqy}$ by $e^{i q \ell \left( \frac{\Pi_x \ell}{\hbar} \right)}$ (and suppress constants) noting specifically that the translation generator $p_x = \Pi_x$ in the Landau gauge we work in; see for instance~\cite{sherafati2016hall}. 

The relevant response functions can be computed at finite frequencies using the Kubo response functions

\begin{align}
    \sigma_H &= \frac{-1}{i \omega^+} \chi_{\left(J_x(q)^\dagger ; J_y (q)\right)} ,\nonumber \\
    \sigma_T &= \frac{-1}{i \omega^+} \chi_{\left(J_x(q)^\dagger ; J_x (q)\right)}, \nonumber \\
    \text{Re} \left[ \eta_H (\omega) \right] &= \frac{1}{\omega} \text{Im} \left[ \chi_{\left(P_{xx} ; P_{xy}\right)} \right] ,
\end{align}

where $\omega^+ = \omega + i0^+$ and $\chi_{A;B}$ is given by

\begin{align}
    \chi_{\left(A;B\right)} = - \sum_{n,m} \rho_n \left[ \frac{\matrixel{n}{A}{m} \matrixel{m}{B}{n}}{\omega - \omega_{mn} + i0^+}  - \frac{\matrixel{n}{B}{m} \matrixel{m}{A}{n}}{\omega + \omega_{mn} + i0^+} \right].
\end{align}

Going forward, we compute these responses at zero temperature, thus the summation on states $n$ will denote states in occupied Landau levels $n = 0, 1, ..., N_L - 1$ while the summation over $m$ will denote states in unoccupied Landau levels $n = N_L, N_L + 1, ...$. It is also useful to define ladder operators via $\Pi_x = (\Pi_+ + \Pi_-)/\sqrt{2}, \Pi_y = (\Pi_+ - \Pi_-)/\sqrt{2}i$ which satisfy $[\Pi_-, \Pi_+] = 1$. These operators cycle between adjacent Landau levels since $[H, \Pi_+] = \hbar \omega_c \Pi_+$ etc. We find
\begin{widetext}
\begin{align}
\text{Re} \left[ \eta_H (\omega) \right] &= \frac{\hbar \omega_c^2}{4 \pi \ell^2} \sum_{m,n} \rho_n \frac{ \abs{\Pi_+^2}_{mn}^2 - \abs{\Pi_-^2}_{mn}^2}{ 4 \omega^2_c - \omega^2}, \nonumber \\
\sigma_{H} (q,\omega) &= -\frac{e^2}{h} \omega_c^2 \sum_{m,n} \rho_n \left[ \frac{\abs{\Pi_+}^2_{mn} - \abs{\Pi_-}^2_{mn}}{\omega^2_c - (\omega^+)^2} + \frac{(q\ell)^2}{2}\frac{\abs{\Pi_+^2}^2_{mn} - \abs{\Pi_-^2}^2_{mn}}{4 \omega^2_c - (\omega^+)^2} - \frac{(q\ell)^2}{8} \frac{\matrixel{m}{O_H}{n}\matrixel{m}{\Pi_+}{n}-n\leftrightarrow m}{ \omega^2_c - (\omega^+)^2} \right], \nonumber \\
O_H &= 3 \{ \Pi_-^2, \Pi_+ \} + 2 \Pi_- \Pi_+ \Pi_-, \nonumber \\
\sigma_{T} (q,\omega) &= -\frac{e^2}{h} \omega_c^2 \sum_{m,n} \rho_n \left[ \frac{i \omega_c}{\omega^+} \frac{\abs{\Pi_+}^2_{mn} - \abs{\Pi_-}^2_{mn}}{ \omega^2_c - (\omega^+)^2} + \frac{(q\ell)^2}{2} \frac{2 i \omega_c}{\omega^+} \frac{\abs{\Pi_+^2}^2_{mn} - \abs{\Pi_-^2}^2_{mn}}{4 \omega^2_c - (\omega^+)^2} - \frac{(q\ell)^2}{2} \frac{i \omega_c}{\omega^+} \frac{\matrixel{m}{O_T}{n}\matrixel{m}{\Pi_+}{n}-n\leftrightarrow m}{ \omega^2_c - (\omega^+)^2} \right], \nonumber \\
O_T &= \{ \Pi_-^2, \Pi_+ \} + \Pi_- \Pi_+ \Pi_-,
\end{align}
\end{widetext}

where $\left|\Pi_\pm\right|^2 \equiv \left|\matrixel{m}{\Pi_\pm}{n}\right|^2$. We now use these to confirm expectations outlined above. In particular, it is evident that

\begin{align}
    \text{Re}[\sigma_T - i \sigma_H] (q,\omega > 0) &= \mathcal{O} \left( q \ell \right)^2 \delta(\omega - \omega_c) \nonumber \\
    \text{Re}[\sigma_T + i \sigma_H] (q,\omega > 0) &= \left( N_L \frac{e^2}{h} \pi \omega_c + \mathcal{O} \left( q \ell \right)^2 \right) \delta(\omega - \omega_c) \nonumber \\    
    &+ \mathcal{O} \left( q \ell \right)^2 \delta ( \omega - 2 \omega_c)
\end{align}

for $N_L$ filled Landau levels at zero temperature. Thus, indeed the magnetic noise differs sharply between the two orientations of the magnetic dipole probe (or for relaxation from the states $\ket{m_s = \pm 1}$ in an NV center). Note that in the limit $ q \rightarrow 0$, the noise vanishes for one orientation while it remains finite for the other, as expected from the classical Drude analysis. 

It is also straightforward to confirm, by examining the matrix elements ,that the Hall viscosity term appears in the real part of the Hall conductivity as expected. In particular,

\begin{align}
    \text{Re}[\sigma_H] (\omega \rightarrow 0^+) = \frac{-N_L e^2}{h} \left[ 1 +  (q \ell)^2 \left( \frac{\eta_H}{n \hbar}  - C \right) \right]
\end{align}

where we used the electron density $n = N_L/(2\pi \ell^2)$, and used $\eta_H$ to represent the zero frequency limit of $\text{Re}[\eta_H (\omega)]$. It is easy to verify that in the limit $\omega \rightarrow 0^+$, one obtains the expected result~\cite{HoyosSon} for the Hall conductivity up to $\mathcal{O} \left(q \ell \right)^2$ for integeral quantum Hall states---$\text{Re}\left[ \sigma_H (q, \omega \rightarrow 0^+) \right] = -\frac{N_L e^2}{h} \left( 1 - \frac{3N_L}{4} (q \ell)^2\right)$ confirming the accuracy of our computations. 

Finally, we can compute the imaginary part of the Hall conducitivity to find

\begin{align}
    \text{Im} \left[ \sigma_H \right] (q, \omega > 0) &= -\frac{N_L e^2}{h} \frac{\pi \omega_c}{2} \big[ \left(1 - (q \ell)^2C \right) \delta(\omega - \omega_c) \nonumber \\
    &+ (q \ell)^2 \frac{\eta_H}{n \hbar} \delta (\omega - 2 \omega_c) \big].
\end{align}

This confirms that the Hall viscosity indeed shows up as a pole at $\omega = 2 \omega_c$ in the dissipative part of $\sigma_H$, and separately from the other piece that is $\mathcal{O} \left(q \ell \right)^2 $. 

We do not know if a similar separation in frequencies applies to interacting fractional quantum Hall states. It would be a useful exercise to compute the real and imaginary parts of the Hall conductivity in these systems to ascertain how applicable these ideas are to the more general setting of fractional quantum Hall states; we leave this to future work. We also note in passing that the same analysis appears to hold for graphene as well where the electrons satisfy the relativistic Dirac equation---this has been pointed out for the real part of the Hall conductivity in Ref.~\cite{sherafati2016hall}. However, it is easy to consider similar computations in the case of graphene and verify that the results found here for the imaginary part of the Hall conductivity hold in graphene as well. We do not provide further details of this computation as it proceeds analogously. 

It is also worth noting that these computations have been performed in the idealized limit of no disorder. We generally expect that disorder will broaden these spectral features, and in general one can measure a combination of the Hall viscosity and the width of the Landau levels by examining the relaxation rate of the probe near $\omega = 2 \omega_c$. This width can be independently ascertained, for instance, using STM measurements. 

\section{TRSB Superconductors}
\label{sec:SCs}

TRSB superconductors are another instance where local probes may contain useful information to detect TRSB phenomenology. A key proponent in this class of supercoductors is the chiral $p$-wave superconductor which is known to support non-Abelian excitations~\cite{ivanov2001non,read1996quasiholes}, specifically Majorana Zero Modes, bound to vortices. Some evidence for time-reversal symmetry breaking in the superconducting state can be found in SrRuO$_4$, UPt$_3$ and URu$_2$Si$_2$ with the help of Kerr rotation measurements~\cite{PhysRevLett.97.167002,schemm2014observation,PhysRevB.91.140506} and muon spin relaxation $\mu$SR measurements~\cite{luke1998time}. However, other experimental data appears to contradict these findings---lack of change of order in SrRuO$_4$ in the presence of uniaxial strain or in-plane field; see for instance~\cite{jerzembeck2024tc,hicks2014strong,taniguchi2015higher}.

A general difficulty in unambiguously ascertaining the nature of the order parameter in this system comes from the multi-orbital nature of superconductivity in this system, the intertwining of spin and orbital degrees of freedom which can substantially alter the interpretation of  TRSB probes such as the Knight shift, and close competition between various even and odd parity superconducting orders, which makes the interpretation of probes that require an external field, harder to fully trust~\cite{mackenzie2017even}. A key advantage of the methodology proposed here is that it is non-invasive. As opposed to NMR based spectroscopic methods, no external probe field is applied on the sample and the signature of TRSB are obtained instead from the spectrum of magnetic fluctuations above a material. 

Another setting purported to exhibit TRSB superconductivity is that of thin flakes of cuprates stacked at twist angles close to $45^\circ$~\cite{can2021high,haenel2022incoherent}. Here Josephson transport measurements have recently been used to provide evidence of TRSB~\cite{zhao2023time}. 

We now detail the computation of the transverse and anomalous Hall conductivity in such chiral superconductors, following the work of Ref.~\cite{lutchynchiralpwave}. The gauge-invariant density and current correlation tensor $K_{\mu \nu} (q)$ is given by 

\begin{align}
    K_{\mu \nu} = Q_{\mu \nu} - \frac{Q_{\mu \rho} q^\rho q^\sigma Q_{\sigma \nu}}{Q_{\alpha \beta} q^\alpha q^\beta}
\end{align}

in terms of non-gauge-invariant correlation tensor $Q_{\mu \nu} (q)$ which is computed from the free electron Green's function with a fixed superconducting order parameter, and $q$ refers to the three-vector $q = (\omega^+, q_x, q_y)$. The expressions for $Q_{\mu \nu}$ are detailed in the appendices of Ref.~\cite{lutchynchiralpwave}. The conductivity tensor can be computed using the relation $\sigma_{ab} (\omega, \vec{q}) = \frac{-K_{ab} (\omega, \vec{q})}{i\omega^+}$ for $a,b \in \{x,y\}$. Here we use these results to compute the relevant parts of the conductivity tensor, that is, $\sigma_T, \sigma_H$. We consider the computation with the following assumptions. First, we assume that the Bogoluibov quasiparticles propagate ballistically in the sample on the relatively small length scale $1/q$; thus, we do not introduce any mechanism for relaxation of electrons and in this way, the results from a straightforward Kubo linear response calculation are appropriate for computing conductivities directly. We also assume that the wave-vector $q \sim 1/z_{\text{NV}}$ is smaller than the Fermi wave-vector $k_F$ in the superconductor and $1/\xi$, where $\xi$ is the coherence length of the fluctuations in the superconducting order parameter, as is appropriate to the results of Ref.~\cite{lutchynchiralpwave}. Next, we compute results to lowest non-zero order in $\omega/qv_F$; this is well justified because $\omega_{\text{NV}} z_{\text{NV}}/v_F \ll 1$ in most scenarios of interest (NV tip distance of $10-100$ nm and $v_F \sim 10^5 - 10^6$ m/s). For simplicity, we also assume an isotropic Fermi surface. Under these assumptions, we arrive at the following result for the transverse conductivity---

\begin{align}
    \text{Re} & \left[ \sigma_T (q, \omega) \right] \approx -\frac{1}{\omega} \text{Im}\left[Q_{xx} (q \hat{y}, \omega)\right] \nonumber \\
    &= \frac{ne^2}{m_e} \frac{1 }{\pi v_F q} \int_{\Delta \sqrt{1 + \left(\frac{\omega}{v_F q}\right)^2}}^\infty dE \; \frac{E^2}{E^2 - \Delta^2} \left( - \pd{f}{E} \right) \nonumber \\
    & \times \int d\theta_p \; \cos^2 \theta_p \; \delta \left( \sin \theta_p - \frac{\omega}{v_F q} \frac{E}{\sqrt{E^2-\Delta^2}}\right),
    \label{eq:sigmaTchiral}
\end{align}

where $n$ is the electron density. We note that the factor of $\left( E/\sqrt{E^2-\Delta^2} \right)^2$ corresponds to the density of states of the Bogoluibov quasiparticles at energy $E, E+\hbar \omega \approx E$. The integral over the angle $\theta_p$ enforces energy conservation in the scattering process in which a Bogoluibov quasiparticle absorbs a photon of wavevector $q \hat{y}$ and energy $\hbar \omega$, and the Fermi factor $-\pd{f}{E}$ ensures that such a process is feasible in the presence of Pauli blocking. Note that the temperature dependence comes from both the Fermi factor, but also the gap $\Delta$ itself. The above result stands for an isotropic gap $\Delta$. However, if we are interested only in low wave-vectors $q k_F \ll 1$, then the above result can be used to obtain the conductivity for an anisotropic gap (as is appropriate for a conventional d-wave superconductor) by averaging for $\Delta (\theta_k) \sim \cos 2 \theta_k$ as it varies for momenta $\vs{k}$ over the Fermi surface. 

The result is reminiscent of the standard calculation of the Hebel-Slichter peak observed in NMR which corresponds to enhanced relaxation rate of nuclear spins immediately below the superconducting transition temperature $T_c$ as compared to above $T_c$. One can show that here this enhancement factor is approximately given by $\text{log}[qv_F/\omega]$.

\begin{figure}[h]
    \includegraphics[width=0.4\textwidth]{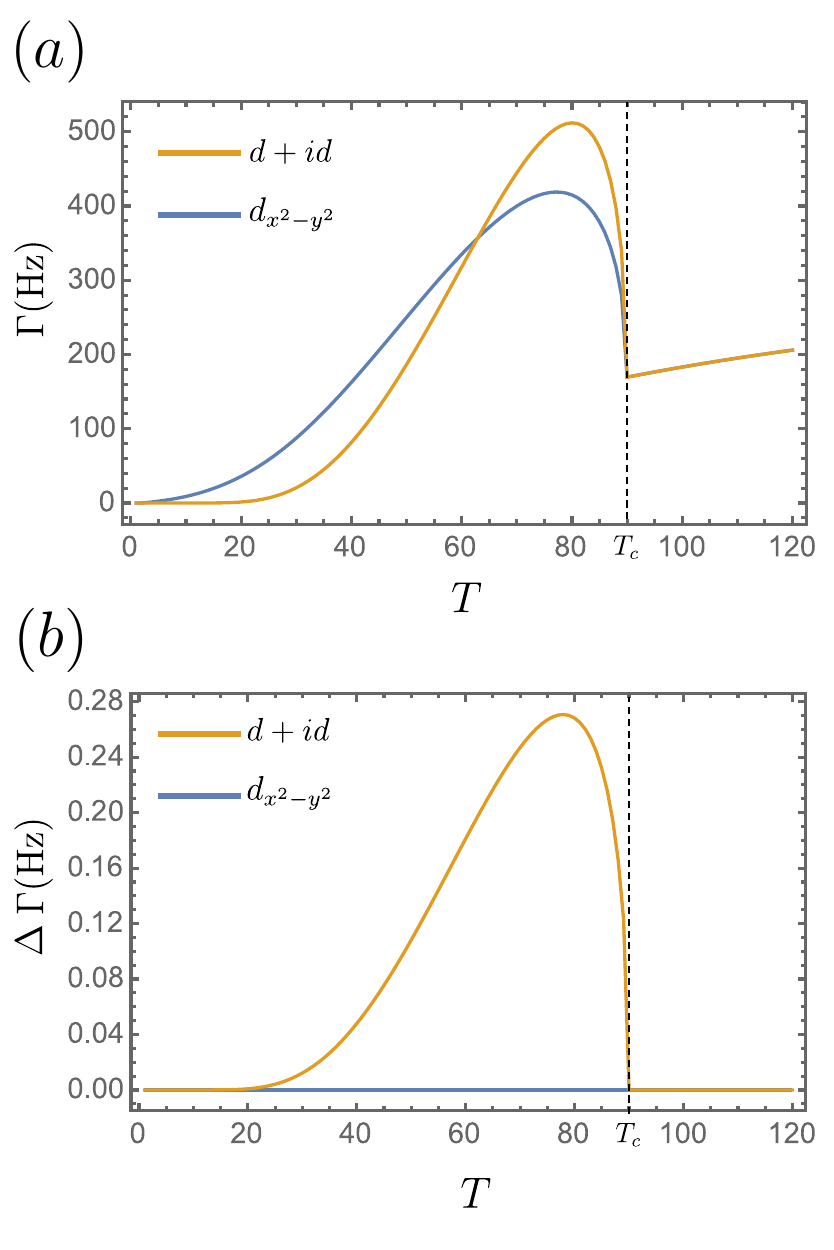}
    \caption{(a) The averaged rate $[\Gamma_{\hat{z}} + \Gamma_{-\hat{z}}]/2$ and (b) difference $\Gamma_{\hat{z}}-\Gamma_{-\hat{z}}$ are shown for a conventional d-wave superconductor (yellow) and a chiral d-wave superconductor (blue) with pairing angular momentum $l = +2$, with $N = 100$ active layers independently (that is, weakly correlated among themselves) contributing to the magnetic noise. We model the parameters on BSCCO thin films where signs of time-reversal symmetry breaking have been observed in flakes transposed upon each other after a 45 degree twist. In particular, we use an electron density $n = 1.0\times 10^{14}/\text{cm}^2$, effective electron mass $m_e = 3 m$ where $m$ is the bare electron mass, an in plane low-frequency dielectric constant $\epsilon = 10$ (which is excepted to be similar to that obtained in the normal phase~\cite{arseev2006theory}; this reduces the differential rate directly by the factor of $\epsilon$) and $T_c = 90$K. The NV to sample distance is assumed to be $z_{\text{NV}} = 20 $nm and assume that the sample thickness is of the same order and contributes . We note that for thin superconducting flakes, the screening may be much weaker and the difference of relaxation rates could be much greater as a consequence. We use $\gamma^2_{\text{NV}} = S(S+1)g^2\mu_B^2 / 2 \hbar^2$, with $S = 1$, $g = 2$ to obtain the relaxation rates from the noise. We also assume a temperature dependence of the gap $\Delta (T)$ given by the usual BCS mean-field result. We note that the differential rate may be larger }
    \label{fig:gammafig}
\end{figure}

We note here that unlike the usual calculation for nuclear spin relaxation, here we are interested in relaxation due to magnetic fluctuations produced by currents and not electronic spins inside the superconductor. Nevertheless, the general result is similar. Another point of difference is that the NV center's relaxation rate is determined by finite (typically low) $q \ll k_F$ fluctuations while nuclear spin relaxation is determined by relaxation at all $q$. This is because nuclear spins can be assumed to be point-like objects residing within the material itself. This has the important implication that the Hebel-Slichter peak is nearly absent in NMR in conventional d-wave superconductors because of the sign changing of the order parameter~\cite{dai2024existence}. Importantly, this enhancement of relaxation below $T_c$ can be observed using NV centers even in conventional d-wave superconductors due to selectivity of low $q$ scattering processes. We visualize this result in terms of the computed relaxation rate of the NV centers in Fig.~\ref{fig:gammafig} (a) for both the chiral and conventional d-wave superconductors, neglecting the Hall contribution to the result. The result for the conventional d-wave case is obtained effectively by averaging the result in Eq.~(\ref{eq:sigmaTchiral}) over superconducting gap drawn from $\Delta (\theta) = \Delta_0 \cos 2 \theta$. Here we see that the noise at lower frequencies in the conventional d-wave case is less suppressed at lower temperatures than the chiral d-wave system due to the presence of gapless quasiparticles at the nodal points. Concomitantly, the Hebel-Slichter peak is suppressed due to a lower pile up of quasiparticle density of states as a consequence of the the gap vanishing at the nodal points. 

Next, for the Hall conductivity, we find

\begin{align}
    \text{Im} & \left[ \sigma_H (q, \omega) \right] \approx -\frac{1}{\omega} \frac{\omega}{q} \text{Re} [ Q^{(a)}_{x0}] \nonumber \\
    &= - \frac{2 \pi l e^2}{h} \frac{\omega}{q v_F} \int_{\Delta \sqrt{1 + \left(\frac{\omega}{v_F q}\right)^2}}^\infty dE \; \frac{\Delta^2}{E^2 - \Delta^2} \left( - \pd{f}{E} \right) \nonumber \\
    & \times \int \frac{d\theta_p}{2\pi} \; \cos^2 \theta_p \; \delta \left( \sin \theta_p - \frac{\omega}{v_F q} \frac{E}{\sqrt{E^2-\Delta^2}}\right). \nonumber \\
\end{align}

We see that the Hall conductivity mimics the temperature, wave-vector and frequency dependencies of the transverse conductivity (in this $\omega \ll qv_F$ regime as discussed above). However, it is directly proportional to $l$, or the pairing angular momentum of the Cooper pairs, and is thus non-zero only when time-reversal symmetry is broken under $T_c$. We also see that it is not proportional to the electronic density in the sample. Observing such a small decay rate in BSCCO is likely to be challenging but within reach based on coherence times of pristine NV centers. Concrete expectation of the relaxation rates as a function of temperature is shown in Fig.~\ref{fig:gammafig} (b), with material properties discussed in the caption.  

We note that in these calculations, we have neglected the contributions to magnetic noise due to phase and amplitude fluctuations of the superconducting order parameter itself which lead to supercurrent fluctuations that can produce a measurable effect close to the superconducting phase transition. The contribution of these fluctuations to the NV relaxation rate for similar device parameters in BSCCO has been computed in Ref.~\cite{liu2025quantum} and appear to explain experimental data. However, for relatively clean samples, where the mean free path of the Bogoluibov quasiparticles reaches tens of nanometers, that is, is larger than $z_{\text{NV}}$, we expect these supercurrent fluctuations to be comparatively weaker than that from quasiparticle scattering as computed here and evidenced in Fig.~\ref{fig:gammafig}.

\section{Discussion and Conclusions}
\label{sec:conclusions}

In this work, we examined the magnetic noise originating from materials with broken time-reversal symmetry. We show that the fluctuation spectrum of the magnetic fields originating from both local magnetic moments and currents in these materials exhibit an asymmetry between opposing chiralities. The relaxation rates of the NV center starting from state $\ket{m_s = 1}$ and $\ket{m_s = -1}$ are sensitive to the opposite chiralities and thus the difference in these relaxation rates isolates time-reversal symmetry breaking phenomena. This distinction is most acute for current noise, where the kernel is such that it connects magnetic fields above a material of a certain chirality to the same chirality of current fluctuations in the material. Thus, in a quantum Hall systems, where one chirality is associated with Landau level excitation, while the other chirality is associated with Landau level de-excitation, this discrepancy is most stark---when the NV center's magnetic dipole starts oriented in the opposite direction as the magnetic field establishing the quantum Hall state, the relaxation rate is zero (modulo ambient electromagnetic fluctuations) at zero temperature. This effect is less pronounced for magnetic fields generated by local magnetic moments because the dipole kernel allows for generation of magnetic fields of both chiralities from fluctuation spectra associated with raising (and lowering) operators, although with different amplitudes.   

A major advantage of such nanoscale probes is that they are non-invasive, in that they do not require application of external fields for measurement of magnetic fluctuations from materials. This can be particularly important when probing materials with several competing orders where the effect of an external field may drastically alter the state of the system. Another advantage of such probes is that they naturally yield local information about current and magnetic susceptibilities which is not obtained from traditional conductance and bulk magnetic susceptibility measurements. In particular, we show that the difference in the relaxation of NV centers with magnetic dipoles oriented either towards or away from the surface of the material, isolates the contribution from current fluctuations proportional to the Hall conductivity at a wave-vector $q \sim 1/z_{\text{NV}}$, where $z_{\text{NV}}$ is the distance of the probe from the surface of the material. Using a range of probes at varying distances from the material, one can then compute the $\mathcal{O} \left( q l \right)^2$ component of the Hall conductivity, which we argue contains information about the Hall viscosity. The Hall viscosity is a sensitive probe of the state of the electronic fluid in the material and is only non-zero in the presence of time-reversal symmetry breaking. It can be used to infer, in particular, the pairing channel of electrons in a chiral superconductor, but also distinguish various candidate fractional quantum Hall states~\cite{read1996quasiholes,PhysRevB.84.085316} at $\nu = 5/2$ which can be understood as paired states of composite fermions but with different angular momenta. 

On a practical note, we point out that for many interesting systems where such probes could provide new insight on TRSB phenomena, the relevant frequencies (for instance), determined by gaps of order $100$ mK to a few K correspond to frequencies $2-100$ GHz which are addressable via NV center relaxometry~\cite{wan2018efficient,hedrich2020parabolic,Xu2023}. The superconducting coherence length $\ell$ is usually in the $10$ nm - $1 \mu$m regime; this implies that for $q \sim 1/z_{\text{NV}} \ell \lesssim 1$, we require the probe-material distance, $z_{\text{NV}}$, to be of a similar scale. This is experimentally feasible. For conductivities of the order of $e^2/h$, one can generally expect the relaxation rate to be about $1$ Hz or even faster if the distance of the NV center is reduced. This also is within the reach of NV center probes set up in cryogenic environments in ultrapure diamond~\cite{abobeih2018one}. We thus anticipate that these probes are indeed well positioned to measure such novel phenomena.  We have provided concrete results for the expected relaxation rate in the case of time-reversal symmetry breaking in stacked twisted BSCCO flakes which have been studied extensively both experimentally and theoretically recently. 


We note that in this work, we hinted at the possibility of inferring the Hall viscosity from the imaginary part of the Hall conductivity. The imaginary part of the Hall conductivity appears naturally in the relaxation time of the NV center because it is the dissipative component associated with the Hall response. A rigorous connection between the \emph{real} part of the Hall conductivity and Hall viscosity was made in Ref.~\cite{HoyosSon}; here we provided some evidence that such a response also manifests itself in the \emph{imaginary} part of the Hall conductivity by considering the appropriate response functions for a the integer quantum Hall phase in a two-dimensional electron gas, and for a chiral superconductor (quoting Ref.~\cite{lutchynchiralpwave}). It would be worthwhile to understand if this connection holds more generally. Alternatively, one can also consider a non-dissipative process such as dephasing of NV centers by magnetic fluctuations. This should probe the \emph{real} part of the finite wavevector Hall conductivity, and is being explored by others~\cite{dolgirev2024}.  

\section{Acknowledgements}
We thank Eugene Demler and Joaquin Rodriguez-Nieva in particular for previous related work along with useful discussions on the current manuscript, as well as Pavel Dolgirev, Ivar Martin, Mike Norman and Thomas Szkopek for useful inputs. KA acknowledges support of the Material Sciences and Engineering Division, Basic Energy Sciences, Office of Science,
United States Department of Energy.

\bibliography{TRSBProbebib}

\end{document}